\documentclass[a4,12pt]{article}

\makeatletter
\renewcommand{\ps@plain}{%
        \renewcommand{\@evenhead}{}%
        \renewcommand{\@oddhead}{}%
        \renewcommand{\@evenfoot}{\hfil\small{\textbf\thepage}\hfil}%
        \renewcommand{\@oddfoot}{\@evenfoot}}%

\renewcommand\section{\@startsection {section}{1}{\z@}%
                                   {-3.5ex \@plus -1ex \@minus -.2ex}%
                                   {2.3ex \@plus.2ex}%
                                   {\reset@font\bfseries}}
\renewcommand\subsection{\@startsection{subsection}{2}{\z@}%
                                     {-3.25ex\@plus -1ex \@minus -.2ex}%
                                     {1.5ex \@plus .2ex}%
                                     {\reset@font\itshape}}
\newlength{\capsize}
\renewcommand{\@makecaption}[2]{%
  \vskip\abovecaptionskip
  \sbox\@tempboxa{\small{\bfseries #1}\/: #2}%
  \ifdim \wd\@tempboxa >\capsize
   {\advance\leftskip by 0.1\textwidth \advance\rightskip by 0.1\textwidth
    {\small {\bfseries #1}: #2}\par}
  \else
    \hbox to\hsize{\hfil\box\@tempboxa\hfil}%
  \fi
  \vskip\belowcaptionskip}

\makeatother

\usepackage{amssymb}
\usepackage{latexsym}
\usepackage{graphics}
\usepackage{graphicx}
\usepackage{psfrag}

\newcommand{\T}{\mathcal{T}}
\newcommand{\To}{\mathcal{T}_1}
\newcommand{\Tp}{\mathcal{T}_p}
\newcommand{\Tr}{\mathrm{Tr}}

\newcommand{\tp}{\tilde{\psi}}

\newcommand{\ip}[2]{\langle #1|#2 \rangle}

\newcommand{\Hi}{\mathcal{H}}
\newcommand{\Hl}{\mathcal{H}_{\ell}}
\newcommand{\D}{\mathcal{D}}

\newcommand{\W}{\mathcal{W}}

\newcommand{\w}{w(x)^{\frac{1}{2}}}

\title{\large \bf Approximating Connections in Loop Quantum Gravity}

\author{{\normalsize \bf Matthias Arnsdorf}\thanks{m.arnsdorf@ic.ac.uk}
\\ 
\textit{\normalsize Blackett Laboratory,} 
\\
\textit{\normalsize Imperial College of Science Technology and Medicine,}
\\ 
\textit{\normalsize London SW7 2BZ, United Kingdom.}
}

\date{\normalsize October 22, 1999}

\begin{document}

\maketitle

\begin{abstract}

We discuss the action
of the configuration operators of loop quantum gravity.  In
particular, we derive the generalised eigenbasis for the Wilson loop
operator and show that the transformation between this basis and the
spin-network basis is given by an expansion in terms of Chebyshev
polynomials.

These results are used to construct states which approximate
connections on the background 3-manifold in an analogous way that the
weave states reproduce area and volumes of a given 3-metric.
This should be necessary for the construction of genuine
semi-classical states that are peaked both in the configuration and
momentum variables.

\end{abstract}

\section{Introduction}

One of the main challenges facing non-perturbative (loop) quantum
gravity to date is to show how general relativity is reproduced in the
appropriate classical limit.  Semi-classical states that have been
considered so far are the weaves~\cite{grot97,ashtekar92}, which are
attempts to approximate classical geometries on a background spatial
3-manifold $\Sigma$. More precisely, one requires that given a
3-metric $g_{ab}$ on $\Sigma$, expectation values of areas and volumes
in the weave state are given by the values determined by $g_{ab}$.

However, area and volume operators only depend on the momentum
variables and we expect that to construct genuine semi-classical or
coherent states configuration variables need to be approximated as
well. One state corresponding to flat space that satisfies these
requirements has been constructed
in~\cite{arnsdorf99b}. To investigate such states in more detail and
provide further examples better control of the configuration operators
is required. In loop quantum gravity these are the cylindrical
functions of connections, which essentially determine holonomies on
paths embedded in $\Sigma$.

In this paper we tackle these issues, and construct states which
approximate connections in the same sense that weaves approximate
3-metrics. In particular, we consider the operator operator $\To$,
which is a Wilson loop in the fundamental representation of
SU(2). Given any connection $A$ on $\Sigma$ we show how states can be
constructed that give expectation values of $\To$ corresponding to
$A$. We call these states ``holonomy weaves''.

To achieve this goal we determine the generalised eigenvectors of the
operator $\To$, which can be understood using the following
analogy. It is known that in infinite dimensions 
operators with continuous spectrum will in general
have non-normalisable eigenvectors. This is already evident in
elementary quantum mechanics where the operator $\hat{p}=-id/dx$ has
the improper eigenvectors $e^{ipx}$.  Nevertheless, in
practice we can consider 
physically relevant wave packets or smeared states.
The analogue of an expansion of an arbitrary state in terms of the
momentum eigenvectors is given by the familiar Fourier transform: 
\[
\tilde{f}(x) = \frac{1}{\sqrt{2\pi}}\int_{-\infty}^{\infty} f(p)e^{ipx} dp,
\]
where $f \in \mathcal{L}^2(\Bbb{R})$.

To make the above ideas more rigorous one can introduce the concepts
of Gel'fand triples or rigged Hilbert spaces~\cite{gelfand64}. In this
context the functions $e^{ipx}$ are considered as linear functionals
acting on a subspace $\D \subset \mathcal{L}^2(\Bbb{R})$, where the
action $e^{ipx}[f]$ is just given by the above integral and the
subspace $\D$ is determined by the requirement that the integral is
well-defined. They are generalised eigenvectors in the sense that:
\[
(\hat{p}'e^{ipx})[f] \equiv e^{ipx}[\hat{p}f] = pe^{ipx}[f],
\]
for all $f \in \D$, where $\hat{p}'$ denotes the natural dual action
of $\hat{p}$.

The key to this approach is that an analogous construction is always
possible for any self-adjoint operator. In fact it can be shown that
there is a sense in which every self-adjoint operator has a complete
set of generalised eigenvectors. In this paper we 
apply these results to loop quantum gravity. In particular, 
we will construct the generalised eigenvectors of the $\To$ operator
and show that the analogue of the above Fourier transform is an
expansion in term of Chebyshev polynomials. This lets us transform
between the spin-network basis and the generalised eigenbasis of the
holonomy operator, which as we shall see greatly simplifies our search
for the holonomy weave.

\section{Generalised eigenvectors of $\To$}
\subsection{Preliminaries}

We begin our discussion by summarising the main facts from loop
quantum gravity that will be needed in the following.  In a series of
papers (e.g.~\cite{rovelli90,ashtekar95,baez95}) a rigorous framework for
quantising canonical general relativity in the new
variables~\cite{ashtekar91,barbero94} has been developed. In this
approach the classical phase space is co-ordinatised by
$\mathfrak{su}(2)$ valued connection one-forms $A^i_a$ on a spatial
three manifold\footnote{$\Sigma$ is usually taken to be
compact. For many applications e.g.\ asymptotically flat spaces, the
non-compact case is more interesting. An extension of the above has
been developed in~\cite{arnsdorf99b}.} $\Sigma$  and a conjugate desitised triad
$\tilde{E}^a_i$, which takes values in the dual of the Lie algebra
$\mathfrak{su}(2)$. The spatial index $a$ and the Lie algebra index $i$
will be suppressed in the following.

The quantum theory is given by a Hilbert space $\Hi$ of cylindrical
functions of connections. Cylindrical because they depend on
connections only via their holonomies on finite graphs. More precisely,
given a set of piecewise analytic paths $\{\gamma_1,\ldots,\gamma_n\}$
that form a Graph $\Gamma$ embedded in $\Sigma$ we consider the
space generated by \emph{gauge invariant}\footnote{Gauge invariance
means invariance under SU(2) gauge transformations of the
connection, which is required by the constraints  of
general relativity. We will not be considering the diffeomorphism or
Hamiltonian constraints in this paper.}
 functions of the type:
\[
\Psi_{\Gamma,f}(A) = f(H_{\gamma_1}(A),\ldots,H_{\gamma_n}(A)),
\]  
where $H_{\gamma_i}(A)$ is the holonomy of the connection $A$ along
the  path $\gamma_i$, which takes
values in SU(2) and $f$ is a function from $SU(2)^n$ to $\Bbb{C}$.
Completion of this function space in the appropriate norms gives us
the Hilbert space $\Hi$. It is equipped with the inner product:
\begin{equation} \label{def-ip}
    \ip{\Psi_1}{\Psi_2} = 
    \int_{\mathrm{SU(2)}^n}f^*_1(g_1,\ldots,g_n)f_2(g_1,\ldots,g_n) dg_1\cdots
    dg_n.
\end{equation}
Here we make use of the fact that if the functions $f_1$ and $f_2$
have a different number of arguments, say $f_1:\mathrm{SU(2)}^{m}
\rightarrow \mathbb{C}$ with $m < n$, we can trivially extend $f_1$ to
a function on $\mathrm{SU(2)}^n$, which does not depend on the last
$n-m$ arguments.  It can be shown that $\Hi$ is spanned by the
so-called spin-network functions which are generalisations of the
Wilson loop. We will be making use of this fact later.

Elementary operators on this space are given by the cylindrical
functions, which act multiplicatively and by certain derivations on
them. We will be concerned with the configuration operators, which
capture information about the connection. In particular, we will study
the spectrum of the Wilson loop operator $\To =
\Tr \left[\rho_1(H_{\ell}(A))\right]$, where $\rho_1$ is the
fundamental representation of SU(2) and $\ell$ is a closed loop in
$\Sigma$. The action of $\To$ is given by:
\[
(\To\Psi_{\Gamma,f})(A) =
\Tr\left[\rho_1(H_{\ell}(A))\right]f(H_{\gamma_1}(A),\ldots,H_{\gamma_n}(A)).
\]

In the next section we will be looking for eigenstates of the operator
$\To$. To do this we first  restrict our attention to the action of
$\To$ in the
  subspace $\Hl$ of $\Hi$ given by
cylindrical function based on $\ell$. 
In the final section we will show
how the results obtained can be extended to deal with more general states.

\subsection{Chebyshev Polynomials}

We begin with the observation that the operator $\To$ has no proper
eigenstates, which is to be expected since as a multiplicative
operator its spectrum should be the continuous interval $[-2,2]$.  To
explore this in more detail we note that any gauge invariant
cylindrical function $\Psi(A)=f(H_{\ell}(A)) \in \Hl$ can be expanded
in the spin-network basis with coefficients $\psi[p]$:
\begin{equation}\label{exp}
\Psi = \sum_{p=0}^{\infty} \psi[p] \Tp,
\end{equation}
where $\Tp(A) = \Tr \left[\rho_p(H_{\ell}(A))\right]$ and $\rho_p$
denote the representations of SU(2) in colour notation, i.e.\
$\dim(\rho_p) = p+1$.  Eigenvectors $\Psi_x$ of $\To$ have to satisfy:
\[
\To\Psi_x = x\Psi_x,
\]
where $x \in [-2,2]$.  In colour notation the representation theory of
SU(2) implies that:
\begin{equation}\label{reps} 
\To(A)\Tp(A)= \T_{p+1}(A) + \T_{p-1}(A),
\end{equation}
for all connections\footnote{Equivalently all results in this paper
can be seen as dealing with the theory of functions on SU(2).}
 $A$ and $p \geq 1$.
Using this equality the
eigenvalue equation becomes:
\[
\psi_x[1]\T_0 + \sum_{p=1}^{\infty} (\psi_x[p-1]+\psi_x[p+1])\Tp =
\sum_{p=0}^{\infty} x\psi_x[p]\Tp,
\]
where the $\psi_x[p]$ are the expansion coefficients of the state $\Psi_x$.
Because of the independence of the $\Tp$'s this gives us the
recursion relation:
\begin{eqnarray*}
\psi_x[1] -x\psi_x[0] &=& 0\\
\psi_x[p+1] -x\psi_x[p] + \psi_x[p-1] &=& 0.
\end{eqnarray*}
If we set  choose\footnote{
This freedom is equivalent to a choice of norm.}
 $\psi[0] = 1$ then the above equations define
the modified Chebyshev Polynomials 
 $\psi_x[p] = S_p(x)$, which are related to the more usual Chebyshev
polynomials of the second kind by $S_p(x) = U_p(x/2)$ i.e.:
\begin{eqnarray*}
\psi_x[0] &=& 1 \\
\psi_x[1] &=& x \\
\psi_x[2] &=& x^2-1 \\
\psi_x[3] &=& x^3-2x \\
&\vdots&
\end{eqnarray*} 
This expansion of $\Psi_x$ in terms of the $\psi[p]$ does not converge
for all $x \in [-2,2]$ and hence there are in general no proper
eigenstates of the $\To$ operator.
As explained in the introduction, one way of approaching this problem is
via the Gel'fand triple construction and the use of generalised
eigenvectors\footnote{Since $\To$ acts multiplicatively one might
expect the eigenvectors to be delta function distributions either on
the space of connections or the group SU(2). This is indeed correct
and there is a close relation between these delta functions and the
generalised eigenvectors we are about to construct as will be explored in
future work.
}. To do this  we look for linear functionals $F_x$ on some dense
subspace $\D \subset \Hl$ that satisfy:
\[
(\To' F_x)[\Psi]\equiv F_x[\To \Psi] = xF_x[\Psi],
\]
for all $\Psi \in \D$, where the subset $\D$ will be determined more
precisely later. Using the basis expansion~(\ref{exp}) of $\Psi$ and
the linearity of $F_x$ we obtain:
\[
\psi[0] F_x[\T_1] + \sum_{p=1}^{\infty} \psi[p]\left(F_x[\T_{p-1}]+
F_x[\T_{p+1}]\right)  =
\sum_{p=0}^{\infty} x\psi[p]F_x[\Tp]
\]
Again, because the coefficients $\psi[p]$ are arbitrary this is solved
by the Chebyshev polynomials:
\[
 F_x[\Tp] = S_p(x).
\]
The difference is that now the polynomials $S_p(x)$ define a genuine basis
in the space of functionals on $\D$.
Since the cylindrical function operators all commute amongst each other
we expect that the generalised eigenvectors $F_x$ of $\To$ will also
diagonalise the operators $\T_p$, for all $p$. This is indeed the case
as we show in Appendix A.

In the next section we will discuss how these generalised eigenvectors
are related to the spin-network basis.

\section{Transformation of bases}

The usefulness of generalised eigenvectors stems from the fact that
they allow us to expand states in $\Hl$ in a very intuitive
fashion. Moreover there is a sense in which $\Hl$ is spanned
by the generalised vectors. We now provide the details of the
transformation between the spin-network basis and the generalised
eigenbasis. As we demonstrate later this greatly simplifies
calculations that are otherwise very intractable.

Let us define the function $\tp(x)$ on the closed interval $[-2,2]$:
\begin{equation}\label{expansion1}
\tp(x) \equiv
\w F_x[\Psi]  =
\w \sum_{p=0}^{\infty}\psi[p] S_p(x),
\end{equation}
where $w(x)$ is the weight function:
\[
w(x) = \frac{1}{2\pi}\sqrt{4-x^2}
\]
                   
As the notation implies $\tp(x)$ will be the expansion coefficients
of the state $\Psi$ in the generalised eigenbasis of the operator
$\To$. In particular, we will see that $\tp(x)^2$ gives the
probability of obtaining the value $x$ when making a measurement of
the trace of the holonomy around the loop $\ell$.
Note that for the definition of $\tp(x)$ we need to require that the
series in equation~(\ref{expansion1}) converges for all $x \in
[-2,2]$. The set of all $\Psi \in \Hl$ for which this is true will be
denoted by $\mathcal{P}$. We will come back to this point later.

First let us investigate the inverse of the above transformation, which uses
the orthogonality properties of the Chebyshev polynomials. In
particular we have:
\[
\int_{-2}^{2} w(x) S_n(x)S_m(x)dx = 
\left\{\begin{array}{ll}
	0 & m \neq n\\
	1 & m = n \end{array}
	\right.
\]
Using this we find:
\begin{equation}
\int_{-2}^{2}\w
\tp(x)S_p(x)dx
 = \int_{-2}^{2} w(x) \sum_{k=0}^{\infty} \psi[k]
S_k(x) S_p(x) dx 
= \psi[p]. \label{expansion2}
\end{equation}
This represents the desired expansion of $\Psi$ in terms of generalised
eigenvectors of $\To$ with eigenvalues in the interval $[-2,2]$.
Note that
in the last step we had to  exchange the order of
integration and taking the limit, which is valid if
$\sum_{p=0}^{\infty} \psi[p] S_p(x)$ converges \emph{uniformly} on
$[-2,2]$. Let $\mathcal{U}$ be the set of $\Psi$ with such coefficients.
Even if $\Psi \not \in \mathcal{U}$ it might still be true that the
series in equation~(\ref{expansion1}) with the coefficients given by
equation~(\ref{expansion2}) converges pointwise to the function
$\tp(x)$, i.e.\ that $\tp(x)$ has a convergent Chebyshev expansion.
The functions $\Psi$ with this property will be the set $\D$ that we
need in our Gel'fand triple construction. In general, we have $\mathcal{U}
\subset \D \subset \mathcal{P}$, and the precise determination of $\D$ is to
our knowledge not yet available. In practice, we will have to make sure
that functions we use are of the correct type by checking if the
expansions~(\ref{expansion1}) and~(\ref{expansion2}) are compatible.

There is a general theorem that states that to any operator there is a
complete set of generalised eigenvectors, which means that given any
state there is unique expansion in terms of them. 
 To see this in the present context we express the norm of
$\Psi$ in terms of the coefficients $\tp(x)$:
\begin{eqnarray*}
\sum_{p=0}^{\infty} \psi[p]^2 &=& \sum_{p=0}^{\infty}
\psi[p] \int_{-2}^{2} \w\tp(x)S_p(x)dx\\
&=&  \int_{-2}^{2}
\w\tp(x)\sum_{p=0}^{\infty}\psi[p] S_p(x)dx
\\
&=& \int_{-2}^{2} \tp(x)^2dx,
\end{eqnarray*}
which is Parseval's equation for orthogonal polynomials.  In
particular, we deduce that if $\tp(x) = \tp'(x)$ for all $x \in [-2,2]$,
then these functions will determine the same spin-network
coefficients $\psi[p]$, i.e.\ $\psi[p] = \psi'[p]$ for all $p$.

The main practical benefit of the generalised eigenvectors comes
through the natural expression of operator actions. We show that for
the expectation value of any operator $\hat{A}$ we have:
\[
\ip{\Psi}{\hat{A}\Psi} =
\int_{-2}^{2}
\w \tp(x)F_x[\hat{A}\Psi] dx,
\]
where the $F_x$ are the generalised eigenvectors of $\To$ as
before. 
The above follows from equations~(\ref{expansion2}) and~(\ref{expansion1})
since:
\begin{eqnarray*}
\ip{\Psi}{\hat{A}\Psi} &=& \sum_{p=0}^{\infty}\psi[p] (\hat{A}\Psi)[p]
\\
&=& 
\sum_{p=0}^{\infty} \psi[p] \int_{-2}^{2} w(x) F_x[\hat{A}\Psi] S_p(x)dx
\\
&=& \int_{-2}^{2}
\w \tp(x)F_x[\hat{A}\Psi] dx,
\end{eqnarray*}
where $(\hat{A}\Psi)[p] = \ip{\Tp}{\hat{A}\Psi}$ are the expansion
coefficients of $\hat{A}\Psi$ in the spin-network basis. 

As an application we show how the expectation value of the operator
$\To$ takes on an intuitive form in the generalised basis, which will
be used in the next section. Using the defining equation of the
generalised eigenvectors, $F_x[\To\Psi] = xF_x[\Psi],$ we derive:
\begin{eqnarray*}
\ip{\Psi}{\To\Psi} &=& \int_{-2}^{2} \w \tp(x) \sum_{p=0}^{\infty}
\psi[p] F_x[\To \Tp] dx\\
&=& \int_{-2}^{2} x \tp(x) \w \sum_{p=1}^{\infty} S_p(x) dx
\\
&=& \int_{-2}^{2} x \tp(x)^2 dx,
\end{eqnarray*}
which shows that $\tp(x)$ can be interpreted as a probability
amplitude.

\section{Holonomy weaves}

We make use of the results in the previous section to construct a
state for loop quantum gravity that approximates a given connection on
the spatial manifold $\Sigma$. This is to be seen in analogy to
previous constructions of weaves which approximate areas and volumes
on $\Sigma$ given a background 3-metric.

As in the weave construction
our state is based on a background graph embedded in $\Sigma$, field
excitations will be concentrated on the edges of the graph. 
For
simplicity we consider a graph $\Gamma$, which is just a union of loops
$\ell_i$ embedded in $\Sigma$. This union is finite if $\Sigma$ is compact. 
 Otherwise,
we need to use a modification of the standard approach to loop gravity
such as the one given in~\cite{arnsdorf99b} to make states
well-defined. For ease of exposition we may restrict ourselves to the
compact case in the following. At the end of this section we will
briefly discuss the implications of the choice of graph and what
modifications are possible.

The holonomy weave, which we will denote by $\W$, is the product of
normalised cylindrical functions $\Psi_i(A) = f_i(H_{\ell_i}(A))$ based
on the loop $\ell_i$:
\[
\W(A)= \prod_{i=0}^n \Psi_i(A),
\]
where the product ranges over all loops in $\Gamma$. This state is
characterised  by the requirement that the
expectation value of $\To^i(A) \equiv \Tr\left[H(\ell_i,A))\right]$
for any loop $\ell_i$ in $\Gamma$ is
given by the value we expect given the background connection $A$ that
we wish to approximate, i.e.:
\[
\ip{\W}{\To^i\W} =  \Tr\left[\rho_1(H_{\ell_i}(A))\right],
\]
for all $\ell_i$.

The link to the results of the previous sections comes because the
definition of the inner product~(\ref{def-ip}) implies that the above
expectation value depends only on the state $\Psi_i$ in $\W$. More
precisely we have:
\begin{eqnarray*}
\ip{\W}{\To^i\W} &=& \int_{\mathrm{SU(2)}^n}  \prod_{j=0}^n f_j^*(g_j)
 \To^i(g_i) \prod_{k=0}^n f_k(g_k) dg_0\ldots dg_n
\\
&=& \prod_{j=0}^n \int_{\mathrm{SU(2)}^n} f_j^*(g_j)\To^i(g_i)f_j(g_j)dg_j
\\
&=& \ip{\Psi_i}{\To^i\Psi_i},
\end{eqnarray*}
since all $\Psi_j$ are normalised. 

Hence, to reproduce the holonomies of any connection, to the accuracy
that the graph allows, we need to choose the $\Psi_j$'s in such a way
that they have the desired expectation value of $\To$. Since we can do
this independently for each of the $\Psi_j$ and since these functions
are each based on just one loop we can make use of the results of the
previous section. To appreciate how our task has simplified let us
first see what this problem looks like in the standard
spin-network basis. Using equation~(\ref{reps}) we deduce:
\[
\ip{\Psi}{\To\Psi} = 2\sum_{p=0}^{\infty}\psi[p]\psi[p+1].
\]
Choosing coefficients $\psi[p]$ to reproduce any of the possible
expectation values seems intractable. However, the formula derived in the last
section allows us to write:
\[
\ip{\Psi}{\To\Psi} = \int_{-2}^{2} x\tp(x)^2 dx.
\]
Hence, to obtain an
expectation value $a \in ]-2,2[$ of $\To$ we need to choose a
function $\tp_a(x)$ on $[-2,2]$ such that:
\[
\int_{-2}^{2} x\tp_a(x)^2 dx = a.
\]
This is solved by any function $\tp_a(x)$ that is symmetric about
$a$ and is normalised on the interval $[-2,2]$. For such a function
$\tp_a(x)$ to define a state $\Psi_a \in \Hl$ we need the further
requirements that $\tp_a(x)$ has a convergent Chebyshev expansion (so
that $\Psi_a \in \D$) and
also --- because of $\w$ in equation~(\ref{expansion1}) --- that
$\tp_a(-2) = \tp_a(2) = 0$.  Note that it is impossible to construct a
state with the expectation value $2$ or $-2$. This is to be expected
since states with expectation values corresponding to the boundary of
a closed spectrum are necessarily eigenstates, but as we have seen
these are not normalisable. Nevertheless, it is possible to obtain
expectation values arbitrarily close to the boundary values. This
raises interesting questions of whether flat space can be constructed
as a cylindrical function state.

\begin{figure}
\begin{center}
\begin{tabular}{lr}
\psfrag{y}{\small $\tp_1(x)$}
\psfrag{x}{\small $x$}
\includegraphics*[height=1.7in,keepaspectratio]{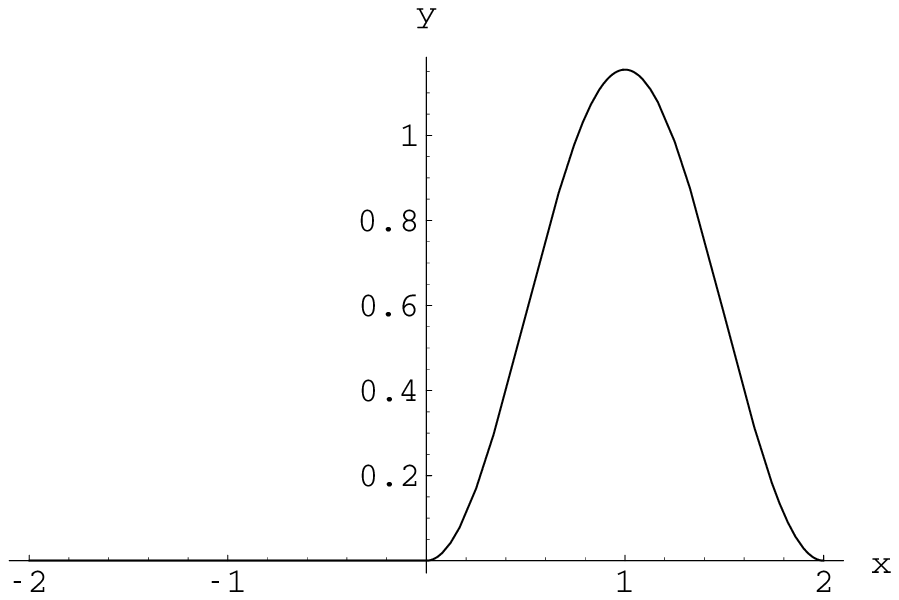}
&
\psfrag{y}{\small $\psi[p]$}
\psfrag{p}{\small $p$}
\includegraphics*[height=1.7in,keepaspectratio]{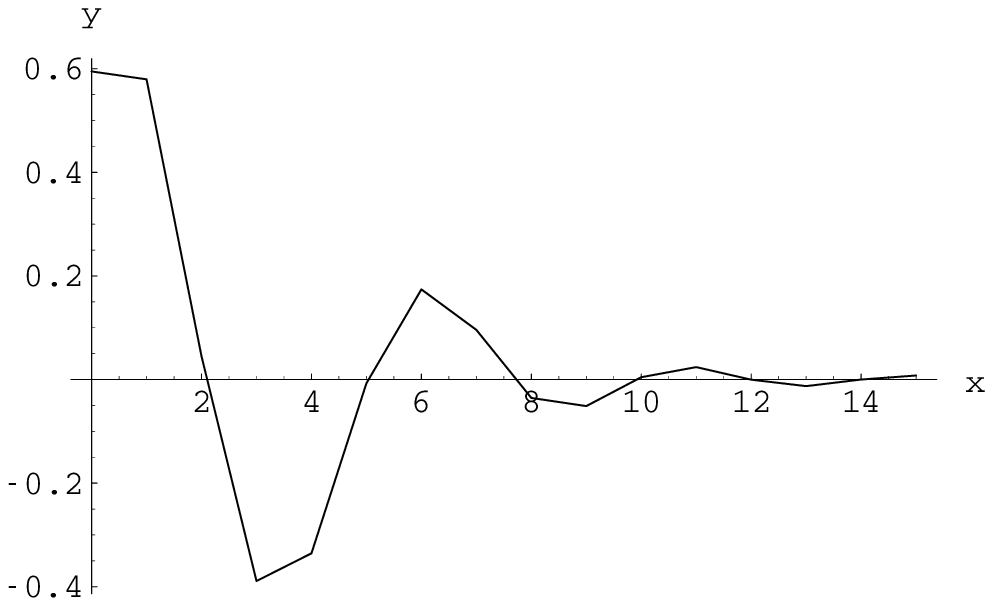}
\end{tabular}
\caption{$\tp_1(x)$ and the normalised spin-network coefficients $\psi[p]$ for
the state $\Psi_1$ in the range $p \in [0,15]$.} \label{spin}
\end{center}
\end{figure}

As an example let us construct a normalised function peaked around $a=1$ to
obtain a state $\Psi_1$ with expectation value 1.
We define:
\[
\tp_1(x) = \left\{ \begin{array}{ll}
		0 &  x \in [-2,0] \\
		\frac{1}{\sqrt{3}}(\cos[\pi(x-1)]+1) & x \in [0,2]
		\end{array} \right.
\]
$\tp_1(x)$ and the corresponding expansion coefficients $\psi[p]$ in
the spin-network basis are shown in figure~\ref{spin}. The $\psi[p]$
fall off rapidly enough to make the transformations between both bases
well-defined.

If weaves are truly to represent semi-classical states then the
requirement that appropriate expectation values are reproduced is not
sufficient. In addition restrictions have to be made on the standard deviations
from the average so that if measurements of areas, volumes, or
holonomies are made on scales large compared to the Planck scale we do
not obtain any deviations from classical values (c.f.~\cite{arnsdorf99b}). 
In the context of the holonomy weaves we note that we can make
deviations around the expectation value of the $\To$ operator
arbitrarily small by sharpening the peak of  the functions
$\tp_a(x)$. 

This is adequate if we happen to measure the holonomy around a loop
that is precisely one of the $\ell_i$ included in $\W$. Since this is
almost never the case we need further requirements on the
nature of the graph underlying the definition of the holonomy weave.
Since excitations are concentrated on the edges of the graph we would
ideally like to cover as much as the manifold $\Sigma$ as possible to
approximate a smooth connection field. But to obtain physically viable
states we have to make sure that these states are weaves in the geometric
sense as well. This places restrictions on density of vertices and
sizes of loops as determined in~\cite{grot97}. In future work we would
like to combine the results from the geometric and the holonomy weaves in
order to construct genuine coherent states.

\section*{Acknowledgements}
I would like to thank Chris Isham for his continued support and
Michael Reisenberger for helpful discussions. 

\section*{Appendix A}

We show that generalised eigenvectors $F_x$ of $\To$ are also generalised
eigenvectors of $\T_p$ with the correct eigenvalues, i.e.:
\[
F_a[\Tp\Psi] = bF_a[\Tp\Psi],
\]
for all $\Psi \in \Hl$, where $a = \To(A)$ and $b=\Tp(A)$ for some
connection $A$. 

First we note the useful relation:
\begin{equation}\label{tp}
\Tp(A) = S_p\left(\To(A)\right).
\end{equation}
This follows from the defining recursion relations for the
polynomials, $S_{p+1}(x) -xS_p(x)+S_{p-1}(x) = 0$, and
equation~(\ref{reps}) by setting $x = \To(A)$.

Next we show that for $p \leq n$:
\[
r_p(x) \equiv \sum_{l=0}^p S_{n-p+2l}(x) = S_p(x)S_n(x)
\]
Consider $r_{p+2} +r_{p}$, where we suppress the $x$ dependence for
notational convenience:
\[
\sum_{l=0}^{p+2} S_{n-(p+2)+2l} + r_p = 2r_p +
S_{n-(p+2)}+  S_{n+p+2}
\]
Using $S_p = xS_{p+1} - S_{p+2}$ we get:
\begin{eqnarray}
r_{p+2} + r_p &=& x(S_{n-(p+1)}+S_{n+p+1}) + S_{n-p} +
S_{n+p} +
2\sum_{l=1}^{p-1} S_{n-p+2l} \label{relation1}
\\
&=& x(S_{n-(p+1)}+S_{n+p+1}) + r_{p} +r_{p-2} \nonumber
\end{eqnarray}  
Repeating these steps we arrive at:
\[
r_{p+2}+r_p = x(S_{n-(p+1)}+\ldots+S_{n-3} +S_{n+3}+\ldots
+S_{n+p+1}) +r_2 +r_0,
\]
for even $p$, and
\[
r_{p+2}+r_p = x(S_{n-(p+1)}+\ldots+S_{n-4} +S_{n+4}+\ldots
+S_{n+p+1}) +r_3 +r_1,
\]
for odd $p$. Now:
\[
r_2 +r_0 = S_{n-2}+S_{n+2} + 2S_n = x(S_{n-1}+S_{n+1}),
\]
using the defining recursion relation for $S$, and:
\[
r_3+r_1 = x(S_{n-2}+S_{n+2})-S_{n-1}-S_{n+1}+2S_{n+1} =
x(S_{n-2}+S_{n}+S_{n+2}),
\]
using equation~(\ref{relation1}).  Hence in both cases we have:
\[
r_{p+2} +r_p =  x\sum_{l=0}^{p+1} S_{n-(p+1)+2l} =xr_{p+1}.
\]
Furthermore we have the initial conditions $r_0
= S_n$ and $r_1 = S_{n-1} + S_{n+1} = xS_n$.
Together with the above this defines the modified Chebyshev polynomials and:
\[
r_p = S_p S_n,
\]
which is the desired result.

Now it is easy to show:
\[
F_a[\Tp\T_n] = bF_a[\T_n],
\] 
for all $n$. Where $a= \To(A)$ is the eigenvalue of the operator $\To$
and $b= S_p(a)$ is the value of $\Tp(A)$ according to
equation~(\ref{tp}). Indeed if $p \leq n$:
\begin{eqnarray*}
F_a[\Tp\T_n] &=& \sum_{l=0}^{p}F_a[\T_{n-p+2l}]
\\
 &=& \sum_{l=0}^{p}S_{n-p+2l}(a) \\
&=& S_p(a)S_n(a) \\
&=& bF_a[\T_n]
\end{eqnarray*}
If $p> n$ we get the same result by symmetry: $\Tp\T_n = \T_n\Tp$.
Hence the the eigenvalue equation is satisfied for all $\T_n$ and
consequently for all $\Psi \in \Hl$ since they are spanned by the
$\T_n \ \Box$.

\bibliography{references}

\end{document}